\documentclass[letterpaper,twoside,twocolumn,english,prb,showpacs,superscriptaddress]{revtex4}
\usepackage[T1]{fontenc}
\usepackage[latin9]{inputenc}
\usepackage{bm}
\usepackage{amsmath}
\usepackage{amssymb}
\usepackage{esint}

\makeatletter

\@ifundefined{textcolor}{}
{%
 \definecolor{BLACK}{gray}{0}
 \definecolor{WHITE}{gray}{1}
 \definecolor{RED}{rgb}{1,0,0}
 \definecolor{GREEN}{rgb}{0,1,0}
 \definecolor{BLUE}{rgb}{0,0,1}
 \definecolor{CYAN}{cmyk}{1,0,0,0}
 \definecolor{MAGENTA}{cmyk}{0,1,0,0}
 \definecolor{YELLOW}{cmyk}{0,0,1,0}
}

\usepackage{babel}

\makeatother

\usepackage{babel}
\begin{document}

\title{Berry Curvature and Phonon Hall Effect}

\author{Tao Qin}

\affiliation{Institute of Physics, Chinese Academy of Sciences, Beijing 100190,
China}

\author{Jianhui Zhou}

\affiliation{Institute of Physics, Chinese Academy of Sciences, Beijing 100190,
China}

\author{Junren Shi}

\affiliation{International Center for Quantum Materials, Peking University, Beijing
100871, China}
\begin{abstract}
We show that an effective magnetic-field acting on phonons naturally
emerges in phonon dynamics of magnetic solids, giving rise to the
phonon Hall effect. A general formula of the intrinsic phonon Hall
conductivity is derived by using the corrected Kubo formula with the
energy magnetization contribution incorporated properly. We thus establish
a direct connection between the phonon Hall effect and the intrinsic
phonon band structure, i.e., phonon Berry curvature and phonon dispersions.
Based on the formalism, we predict that phonon could also have quantum
Hall effect in certain topological phonon systems. In the low temperature
regime, we predict that the phonon Hall conductivity is proportional
to $T^{3}$ for ordinary phonon systems, while that for the topological
phonon system has the linear $T$ dependence with the quantized temperature
coefficient. 
\end{abstract}

\pacs{66.70.-f, 72.20.Pa, 72.10.Bg, 72.15.Gd}

\maketitle

\section{INTRODUCTION}

The phonon Hall effect (PHE) has been discovered recently in $\mathrm{Tb_{3}Ga_{5}O_{12}}$(TGG)~\cite{Strohm2005,Inyushkin2007}:
a magnetized solid could give rise to a temperature difference between
two edges of the sample in the direction transverse to both the magnetization
and the driving thermal flow. The discovery of the Hall effect of
the neutral carriers such as phonon has incited great theoretical
interests~\cite{Sheng2006,Kagan2008,Wang2009,Zhang2010,Agarwalla2011}.
Most of the theories relate the effect to the Raman spin-lattice coupling~\cite{Kronig1928},
and the standard linear response theory, i.e., the Kubo formula~\cite{Sheng2006,Wang2009,Zhang2010,Agarwalla2011}
or its equivalent~\cite{Kagan2008}, are employed to calculate the
thermal Hall coefficient. These investigations, while all focusing
on the intrinsic limit, have not yet revealed a simple connection
between the phonon Hall effect and the intrinsic phonon band structure
for a general system, as that done in electron systems~\cite{Jungwirth2002}.
There is also a natural and interesting question: could phonon systems
have quantum (anomalous) Hall effect as well~\cite{Prodan2009,Zhang2010,Zhang2011}?

In this paper, we establish a direct connection between the PHE and
the intrinsic phonon band structure, i.e., phonon Berry curvature
and phonon dispersions. To do this, first, we derive the general phonon
dynamics applicable for magnetic solids, incorporating the Mead-Truhlar
term in the Born-Oppenheimer approximation~\cite{Mead1979}. The
resulting dynamics contains an effective magnetic field acting on
phonons, which gives rise to the PHE. It clarifies the microscopic
origin of the spin-lattice coupling, and is readily amendable for
the first principles calculation. Second, we calculate the thermal
Hall coefficient of the system using the corrected Kubo formula, incorporating
the contribution of the energy magnetization~\cite{Qin2011,Cooper1997},
which is overlooked in all previous calculations. As the result, we
obtain a general formula for calculating the intrinsic phonon Hall
coefficient. Based on the formalism, we predict that phonon could
also have quantum (anomalous) Hall effect in properly defined topological
phonon systems. In the low temperature regime, we predict that the
phonon Hall conductivity is proportional to $T^{3}$ for the ordinary
phonon systems, while that for the topological phonon systems has
the linear $T$ dependence with the quantized temperature coefficient.

This paper is organized as follow. In Sec.~II, we discussed the general
phonon dynamics in the magnetic field. In Sec.~III, we presented
our central formula for the phonon Hall conductivity. In Sec.~IV,
we gave the definition for the topological phonon systems and also
discussed the possibility to realize it. In Sec.~V, we show the $T^{3}$
law in the low temperature limit for the phonon Hall conductivity.
In Sec.~VI, we have a brief summary. We also include an Appendix
section to show details of derivations.

\section{PHONON DYNAMICS OF MAGNETIC SOLIDS}

\noindent The Hamiltonian for the nucleus motion of a magnetic solid
is determined by the complete form of the Born-Oppenheimer approximation~\cite{Mead1979}:
\begin{equation}
\hat{H}=\sum_{l\kappa}\frac{\left(-\mathrm{i}\hbar\boldsymbol{\nabla}_{l\kappa}-\boldsymbol{A}_{l\kappa}\left(\left\{ \boldsymbol{R}\right\} \right)\right)^{2}}{2M_{\kappa}}+V_{\mathrm{eff}}\left(\boldsymbol{R}\right)\,,\label{eq:realh}
\end{equation}
where $\bm{\nabla}_{l\kappa}=\partial/\partial\bm{u}_{l\kappa}$,
and $V_{\mathrm{eff}}\left(\left\{ \boldsymbol{R}\right\} \right)=\tilde{E}_{\mathrm{e}}\left(\left\{ \boldsymbol{R}\right\} \right)+E_{\mathrm{i}}\left(\left\{ \boldsymbol{R}\right\} \right)$
is the effective interacting potential for nuclei, including the direct
Coulomb interaction between nuclei $E_{\mathrm{i}}\left(\left\{ \boldsymbol{R}\right\} \right)$,
as well as the nucleus interaction mediated by electrons: $\tilde{E}_{\mathrm{e}}\left(\left\{ \boldsymbol{R}\right\} \right)=E_{0}\left(\left\{ \boldsymbol{R}\right\} \right)+(\hbar^{2}/2M_{\kappa})\sum_{l\kappa}\left[\left\langle \boldsymbol{\nabla}_{l\kappa}\Phi_{0}|\boldsymbol{\nabla}_{l\kappa}\Phi_{0}\right\rangle -\left|\left\langle \Phi_{0}|\boldsymbol{\nabla}_{l\kappa}\Phi_{0}\right\rangle \right|^{2}\right]$,
where $E_{0}\left(\left\{ \boldsymbol{R}\right\} \right)$ is the
energy of the ground state $\left|\Phi_{0}(\{\bm{R}\})\right\rangle $
of the electron subsystem at the instantaneous nucleus positions $\left\{ \boldsymbol{R}\right\} $.
For the crystalline solid, the equilibrium positions of nuclei form
a Bravais lattice, and we use $\left\{ \boldsymbol{R}\right\} \equiv\left\{ \boldsymbol{R}_{l\kappa}^{0}+\boldsymbol{u}_{l\kappa},\, l=1\dots N;\,\kappa=1\dots r\right\} $
with $\bm{R}_{l\kappa}^{0}\equiv\bm{R}_{l}^{0}+\bm{d}_{\kappa}$,
where $\boldsymbol{R}_{l}^{0}$ is the center equilibrium position
of $l$-th unit cell, $\boldsymbol{d}_{\kappa}$ is the equilibrium
position of $\kappa$-th nucleus (with mass $M_{\kappa}$) related
to the center, $\boldsymbol{u}_{l\kappa}$ is the corresponding vibration
displacement, $N$ denotes the total number of the unit cells, and
$r$ is the number of atoms in each unit cell.

The most notable feature of Eq.~\eqref{eq:realh} is the presence
of the vector potential $\boldsymbol{A}_{l\kappa}\left(\left\{ \boldsymbol{R}\right\} \right)\equiv\mathrm{i}\hbar\left\langle \Phi_{0}\left(\left\{ \boldsymbol{R}\right\} \right)|\boldsymbol{\nabla}_{l\kappa}\Phi_{0}\left(\left\{ \boldsymbol{R}\right\} \right)\right\rangle $
($\boldsymbol{\nabla}_{l\kappa}\equiv\partial/\partial\boldsymbol{u}_{l\kappa}$),
as first pointed out by Mead and Truhlar~\cite{Mead1979}. In the
modern language, the vector potential is related to the Berry phase~\cite{Jackiw1988}.
The corresponding ``physical field'' at the limit $\bm{u}_{l\kappa}\rightarrow0$
is: 
\begin{equation}
G_{\alpha\beta}^{\kappa\kappa^{\prime}}(\bm{R}_{l}^{0}-\bm{R}_{l^{\prime}}^{0})=2\hbar\mathrm{Im}\left.\left\langle \frac{\partial\Phi_{0}}{\partial u_{\beta,l^{\prime}\kappa^{\prime}}}\right|\left.\frac{\partial\Phi_{0}}{\partial u_{\alpha,l\kappa}}\right\rangle \right|_{\bm{u}_{l\kappa}\rightarrow0}\,,\label{eq:GRealSpace}
\end{equation}
where $\alpha,\beta=x,y,z$, and the translational symmetry dictates
that it must be a function of $\bm{R}_{l}^{0}-\bm{R}_{l^{\prime}}^{0}$.

In the external magnetic field, the ion will experience two vector
potentials~\cite{Resta2000}: one from the real magnetic field and
the other from the Berry phase. The role of the latter is crucial
here. In a single atom, there is exact cancellation between the two,
or the neutral atom would be deflected in the magnetic field. Therefore,
it is not right to deal with the ion in the magnetic field directly
by the ``minimal substitution''~\cite{Holz1972}. In the lattice
system, the cancellation is not exact, and the spin-orbit coupling
of the electron will give rise to nonzero $G_{\alpha\beta}^{\kappa\kappa^{\prime}}\left(\bm{R}_{l}^{0}-\bm{R}_{l^{\prime}}^{0}\right)$.

To proceed, we adopt the usual harmonic approximation by expanding
$V_{\mathrm{eff}}\left(\left\{ \boldsymbol{R}\right\} \right)$ to
the second order of the vibration displacement $\boldsymbol{u}_{l\kappa}$~\cite{Callaway}.
For the periodic lattice, it is more convenient to use the Fourier-transformed
displacement $\boldsymbol{u}_{\kappa\bm{k}}=\sqrt{M_{\kappa}/N}\sum_{l}\bm{u}_{l\kappa}e^{-\mathrm{i}\boldsymbol{k}\cdot\boldsymbol{R}_{l\kappa}^{0}}$
with $\boldsymbol{k}$ being the quasi-momentum, and the Hamiltonian
can be written as: 
\begin{align}
\hat{H} & =\frac{1}{2}\sum_{\boldsymbol{k}}\left(\hat{P}_{\boldsymbol{k}}^{\dagger}\hat{P}_{\boldsymbol{k}}+\hat{u}_{\boldsymbol{k}}^{\dagger}D_{\boldsymbol{k}}\hat{u}_{\boldsymbol{k}}\right)\,,\label{eq:fourierH}
\end{align}
where $\hat{P}_{\boldsymbol{k}}$ and $\hat{u}_{\boldsymbol{k}}$
have $3r$ components with $\hat{P}_{\kappa\alpha,\boldsymbol{k}}\equiv-\mathrm{i}\hbar\partial/\partial u_{\kappa\alpha,-\boldsymbol{k}}-A_{\kappa\alpha,\boldsymbol{k}}$
and $\hat{u}_{\kappa\alpha,\boldsymbol{k}}$, $\alpha=x,y,z$ respectively,
$A_{\kappa\alpha,\boldsymbol{k}}=\mathrm{i}\hbar\left\langle \Phi_{0}|\partial\Phi_{0}/\partial u_{\kappa\alpha,-\boldsymbol{k}}\right\rangle $,
and $D_{\boldsymbol{k}}$ is the $3r\times3r$ dynamical matrix~\cite{Callaway}.

The momentum $\hat{P}_{\kappa\alpha,\boldsymbol{k}}$ has the commutation
relation at the limit $\bm{u}_{l\kappa}\rightarrow0$: 
\begin{align}
\left[\hat{P}_{\kappa\alpha,\boldsymbol{k}},\,\hat{P}_{\kappa^{\prime}\beta,\boldsymbol{k}^{\prime}}^{\dagger}\right] & =\mathrm{i}\hbar G_{\kappa\alpha,\kappa^{\prime}\beta}\left(\boldsymbol{k}\right)\delta_{\boldsymbol{k}\boldsymbol{k}^{\prime}}\,,
\end{align}

\begin{equation}
G_{\kappa\alpha,\kappa^{\prime}\beta}\left(\bm{k}\right)=\frac{1}{\sqrt{M_{\kappa}M_{\kappa^{\prime}}}}\sum_{l}G_{\alpha\beta}^{\kappa\kappa^{\prime}}(R_{l}^{0})e^{-\mathrm{i}\bm{k}\cdot(\bm{R}_{l}^{0}+\bm{d}_{\kappa\kappa^{\prime}})}\,,\label{eq:Gk}
\end{equation}
where $\bm{d}_{\kappa\kappa^{\prime}}\equiv\bm{d}_{\kappa}-\bm{d}_{\kappa^{\prime}}$.
$G_{\kappa\alpha,\kappa^{\prime}\beta}(\boldsymbol{k})$ acts like
an effective magnetic field for the phonon dynamics. Using Eq.~(\ref{eq:GRealSpace})
and (\ref{eq:Gk}), the quantity is readily calculable for the real
materials using the first principles approach.

The general phonon dynamics for a magnetic solid can now be determined.
From Eq.~\eqref{eq:fourierH}, the linearized canonical equations
of motion read:

\begin{align}
\dot{\hat{u}}_{\boldsymbol{k}} & =\hat{P}_{\boldsymbol{k}}\,,\\
\dot{\hat{P}}_{\boldsymbol{k}} & =-D_{\boldsymbol{k}}\hat{u}_{\boldsymbol{k}}+G_{\boldsymbol{k}}\hat{P}_{\boldsymbol{k}}\,,
\end{align}
where $G_{\boldsymbol{k}}$ is a $3r\times3r$ matrix with the component
$G_{\kappa\alpha,\nu\beta}(\boldsymbol{k})$.

The corresponding eigen-equation is: 
\begin{align}
\omega_{\boldsymbol{k}i}\psi_{\bm{k}i} & =\left[\begin{array}{cc}
0 & \mathrm{i}\\
-\mathrm{i}D_{\boldsymbol{k}} & \mathrm{i}G_{\boldsymbol{k}}
\end{array}\right]\psi_{\bm{k}i}\equiv\tilde{H}_{\boldsymbol{k}}\psi_{\bm{k}i}\,,\label{eq:matrix}
\end{align}
where $\psi_{\bm{k}i}$ is the $i$th eigen solution, and $\omega_{\bm{k}i}$
is the corresponding eigen frequency. For the non-hermitian $\tilde{H}_{\bm{k}}$,
we define $\bar{\psi}_{\bm{k}i}=\psi_{\bm{k}i}^{\dagger}\tilde{D}_{\bm{k}}$
with $\tilde{D}_{\bm{k}}\equiv\left[\begin{array}{cc}
D_{\bm{k}} & 0\\
0 & I_{3r\times3r}
\end{array}\right]$ and $I_{3r\times3r}$ is unit matrix of dimension $3r\times3r$,
and $\psi_{\bm{k}i}$ is normalized by $\bar{\psi}_{\bm{k}i}\psi_{\bm{k}j}=\delta_{ij}$.
Note that the $6r$ eigen-solutions can be divided into two groups
of $3r$ positive and negative energy branches, and the two are related
by $\omega_{\boldsymbol{k}i}^{(-)}=-\omega_{-\boldsymbol{k}i}^{(+)}$
and $\psi_{\bm{k}i}^{\left(-\right)}=\psi_{-\bm{k}i}^{\left(+\right)\ast}$
with $1\leq i\leq3r$. This is a result of the symmetries $G_{\boldsymbol{k}}^{*}=G_{-\boldsymbol{k}}$
and $D_{\boldsymbol{k}}^{*}=D_{-\boldsymbol{k}}$. In the following
we would only use the positive energy branches and drop the superscript
$\left(+\right)$ for brevity.

The Hamiltonian Eq.~(\ref{eq:fourierH}) can be diagonalized with
the basis $\psi_{\bm{k}}$. By defining the field operator $\hat{\Psi}_{\bm{k}}\equiv(\hat{u}_{\bm{k}},\hat{P}_{\bm{k}})^{T}$,
Eq.~\eqref{eq:fourierH} can be rewritten as $\hat{H}=(1/2)\sum_{\bm{k}}\hat{\bar{\Psi}}_{\bm{k}}\hat{\Psi}_{\bm{k}}$
with $\hat{\bar{\Psi}}_{\bm{k}}\equiv\hat{\Psi}_{\bm{k}}^{\dagger}\tilde{D}_{\bm{k}}$.
Introducing the transformation

\begin{equation}
\hat{\Psi}_{\bm{k}}=\sum_{i=1}^{3r}\sqrt{\hbar\omega_{\bm{k}i}}\psi_{\bm{k}i}\hat{a}_{\bm{k}i}+\sqrt{\hbar\omega_{-\bm{k}i}}\psi_{-\bm{k}i}^{\ast}\hat{a}_{-\bm{k}i}^{\dagger}\,,\label{eq:transform}
\end{equation}
with $\left[\hat{a}_{\bm{k}i},\hat{a}_{\bm{k}j}^{\dagger}\right]=\delta_{ij}$,
$i,j=1\dots3r$, we can recover all the commutation relations, and
diagonalize the Hamiltonian to: 
\begin{equation}
\hat{H}=\sum_{\boldsymbol{k};i=1}^{3r}\hbar\omega_{\boldsymbol{k}i}\left(\hat{a}_{\boldsymbol{k}i}^{\dagger}\hat{a}_{\boldsymbol{k}i}+\frac{1}{2}\right)\,.\label{eq:phonon}
\end{equation}

Similar to the electronic dynamics in magnetic solids~\cite{Niu1999},
the intrinsic phonon band structure is determined not only by the
phonon dispersions, but also by the Berry connections of the phonon
bands. We can define the phonon Berry connection as $\bm{\mathcal{A}}_{\bm{k}i}\equiv\mathrm{i}\bar{\psi}_{\bm{k}i}\left(\partial\psi_{\bm{k}i}/\partial\bm{k}\right)$,
and the corresponding Berry curvature as, 
\begin{equation}
\bm{\Omega}_{\bm{k}i}=-\mathrm{Im}\left[\frac{\partial\bar{\psi}_{\bm{k}i}}{\partial\bm{k}}\times\frac{\partial\psi_{\bm{k}i}}{\partial\bm{k}}\right]\,.
\end{equation}
We will show that the phonon Berry curvatures $\bm{\Omega}_{\bm{k}i}$
and the phonon dispersions $\omega_{\bm{k}i}$ will fully determine
the intrinsic phonon Hall conductivity. On the other hand, the inter-band
Berry curvatures proposed in some of previous studies are not needed
in general~\cite{Zhang2010}.

\section{PHONON HALL CONDUCTIVITY}

\noindent Following the established general procedure~\cite{Qin2011},
we can calculate the thermal Hall coefficient contributed by phonons.
For a magnetic system, the transport thermal Hall coefficient includes
two parts of contributions: the usual linear response contribution
$\kappa^{\mathrm{Kubo}}$ and the contribution from the energy magnetization
$\bm{M}_{E}$: 
\begin{equation}
\kappa_{xy}^{\mathrm{tr}}=\kappa_{xy}^{\mathrm{Kubo}}+\frac{2M_{E}^{z}}{TV}\,,\label{eq:kappatr}
\end{equation}
where $V$ is the total volume of the system, and $T$ is the temperature.
$\bm{M}_{E}$ is the circulation of the phonon energy current, and
the reason of the circulation can only be attributed to our effective
magnetic field.

\subsection{Kubo contribution $\kappa_{xy}^{Kubo}$}

$\kappa_{xy}^{\mathrm{Kubo}}$ is determined by the usual Kubo formula~\cite{Luttinger1964}:
\begin{equation}
\kappa_{xy}^{\mathrm{Kubo}}=\frac{1}{Vk_{B}T^{2}}\lim_{s\rightarrow0}\lim_{\bm{q}\rightarrow0}\int_{0}^{\infty}\mathrm{d}te^{-st}\left\langle \hat{J}_{E,-\bm{q}}^{y};\,\hat{J}_{E,\bm{q}}^{x}(t)\right\rangle \,,
\end{equation}

\noindent where $\hat{\bm{J}}_{E,\bm{q}}$ is the Fourier transformed
energy current operator in the wave-vector $\bm{q}$, and $\left\langle ;\right\rangle $
denotes the Kubo canonical correlation~\cite{Kubo1983}.

Using the procedure developed by Hardy~\cite{Hardy1963}, and with
the harmonic approximation and in the small $\bm{q}$ limit, we find
that,~{[}See Appendix A1{]}: 
\begin{align}
\hat{\bm{J}}_{E,\bm{q}} & =\frac{1}{8}\sum_{\bm{k}}\hat{\bar{\Psi}}_{\bm{k}}\left(\tilde{\bm{V}}_{\bm{k}}+\tilde{\bm{V}}_{\bm{k}+\bm{q}}\right)\hat{\Psi}_{\bm{k}+\bm{q}}+(\mathrm{h.c.},\bm{q}\rightarrow-\bm{q})\,,\label{eq:current}
\end{align}
where $\tilde{\bm{V}}_{\bm{k}}\equiv\bm{\nabla}_{\bm{k}}\tilde{H}_{\bm{k}}$,
and $(\mathrm{h.c.},\bm{q}\rightarrow-\bm{q})$ denotes the hermitian
conjugation of the first term after replacing $\bm{q}$ by $-\bm{q}$.

With the phonon current operator $\hat{\bm{J}}_{E,\bm{q}}$, we have
{[}See Appendix B{]}, 
\begin{equation}
\kappa_{xy}^{\mathrm{Kubo}}=\frac{\hbar}{VT}\sum_{\bm{k};i=1}^{3r}\mathcal{M}_{\bm{k}i}^{z}\omega_{\bm{k}i}\left(n_{\bm{k}i}+\frac{1}{2}\right)\,,\label{eq:kuboxy}
\end{equation}
where $\bm{\mathcal{M}}_{\bm{k}i}=\mathrm{Im}\left[\frac{\partial\bar{\psi}_{\bm{k}i}}{\partial\bm{k}}\times\tilde{H}_{\bm{k}}\frac{\partial\psi_{\bm{k}i}}{\partial\bm{k}}\right]$,
and $n_{\bm{k}i}\equiv n_{B}(\hbar\omega_{\bm{k}i})$ is the Bose-Einstein
distribution. The presence of the ``zero point'' contribution (the
extra $\frac{1}{2}$ inside the parenthesis) is due to the phonon
number non-conserving terms (e.g., $a_{\bm{k}i}a_{\bm{k}+\bm{q}j}$
or $a_{\bm{k}i}^{\dagger}a_{\bm{k}+\bm{q}j}^{\dagger}$) in the energy
current operator, which were often improperly dropped in many previous
calculations~\cite{Sheng2006,Kagan2008,Wang2009}. As a result of
the zero point contribution, $\kappa_{xy}^{\mathrm{Kubo}}$ diverges
when $T\rightarrow0$.

\subsection{Phonon energy magnetization $M_{E}^{z}$}

The unphysical divergence can be removed by the second term of Eq.~\eqref{eq:kappatr}.
The energy magnetization is determined as follows~\cite{Qin2011}:
\begin{align}
 & 2M_{E}^{z}-T\frac{\partial M_{E}^{z}}{\partial T}=\tilde{M}_{E}^{z}\,,\label{eq:MEz}\\
 & \tilde{M}_{E}^{z}=\frac{1}{k_{B}T}\left.\mathrm{i}\frac{\partial}{\partial q_{y}}\left\langle \hat{h}_{\bm{-q}};\,\hat{J}_{E,\bm{q}}^{x}\right\rangle \right|_{\bm{q}\rightarrow0}\,,\label{eq:MEtilt}
\end{align}
where $h_{\bm{q}}$ is the Fourier transformed energy density operator:
\begin{align}
\hat{h}_{\bm{q}} & =\frac{1}{2}\sum_{\bm{k}}\hat{\bar{\Psi}}_{\bm{k}}\hat{\Psi}_{\bm{k}+\bm{q}}\,.\label{eq:hq}
\end{align}
The energy current operator defined in Eq.~\eqref{eq:current} should
satisfy the scaling law necessary for the applicability~\cite{Qin2011}
of Eqs.~(\ref{eq:MEz}--\ref{eq:MEtilt}) {[}See Appendix A2{]}. 

For the phonon energy magnetization in Eqs.~(\ref{eq:MEz}--\ref{eq:MEtilt}),
we have {[}See Appendix C{]}:

\begin{multline}
\tilde{M}_{E}^{z}=-\frac{\hbar}{2}\sum_{\boldsymbol{k};i=1}^{3r}\omega_{\bm{k}i}\biggl[\Omega_{\bm{k}i}^{z}\omega_{\bm{k}i}^{2}n_{\bm{k}i}^{\prime}\\
+\mathcal{M}_{\bm{k}i}^{z}\biggl(2n_{\bm{k}i}+\omega_{\bm{k}i}n_{\bm{k}i}^{\prime}+1\biggr)\biggr]\,,\label{eq:PEM}
\end{multline}
where $n_{\bm{k}i}^{\prime}=\partial n_{\bm{k}i}/\partial\omega_{\bm{k}i}$.
$M_{E}^{z}$ is obtained by integrating over the temperature $T$
with the boundary condition that $2M_{E}^{z}$ coincides with $\tilde{M}_{E}^{z}$
when $T=0$.

\subsection{Phonon Hall conductivity}

We are on a position to present our central result for the phonon
Hall conductivity. 
\begin{equation}
\kappa_{xy}^{\mathrm{tr}}=-\frac{(\pi k_{B})^{2}}{3h}Z_{\mathrm{ph}}T-\frac{1}{T}\int d\epsilon\epsilon^{2}\sigma_{xy}\left(\epsilon\right)\frac{dn\left(\epsilon\right)}{d\epsilon}\,,\label{eq:kappaxytr}
\end{equation}
where 
\begin{equation}
\sigma_{xy}\left(\epsilon\right)=-\frac{1}{V\hbar}\sum_{\hbar\omega_{\bm{k}i}\leq\epsilon}\Omega_{\bm{k}i}^{z}\,,\label{eq:phononconc}
\end{equation}
and 
\begin{equation}
Z_{\mathrm{ph}}=\frac{2\pi}{V}\sum_{\bm{k};i=1}^{3r}\Omega_{\bm{k}i}^{z}\,.
\end{equation}
Equation~\eqref{eq:kappaxytr} gives the general formula of the intrinsic
phonon Hall conductivity for a magnetic solid. As expected, the intrinsic
phonon Hall conductivity is fully determined by the dispersions and
the Berry curvatures.

In the following, we show how the topological term emerges naturally
in Eq.~\eqref{eq:kappaxytr}.

First, we express $\kappa_{xy}^{\mathrm{Kubo}}$, $\tilde{M}_{E}^{z,\mathrm{inter}}$
and $\tilde{M}_{E}^{z,\mathrm{intra}}$ as: 
\begin{align}
\kappa_{xy}^{\mathrm{Kubo}}= & \frac{1}{2T}\int d\epsilon\epsilon m_{1z}^{6r}\left(\epsilon\right)n\left(\epsilon\right)\,,\\
\tilde{M}_{E}^{z,\mathrm{inter}}= & -\frac{1}{2}\int d\epsilon\epsilon m_{1z}^{6r}\left(\epsilon\right)n\left(\epsilon\right)\,,\\
\tilde{M}_{E}^{z,\mathrm{intra}}= & -\frac{1}{4}\int d\epsilon\left(m_{1z}^{6r}\left(\epsilon\right)-\frac{1}{\hbar}\epsilon\tilde{\sigma}_{xy}^{6r}\left(\omega\right)\right)\epsilon^{2}\frac{\partial n\left(\epsilon\right)}{\partial\epsilon}\,,
\end{align}
where 
\begin{align}
m_{1z}^{6r}\left(\epsilon\right)= & \frac{1}{V}\sum_{\boldsymbol{k},i=1}^{6r}\mathcal{M}_{\bm{k}i}^{z}\delta\left(\epsilon-\hbar\omega_{\bm{k}i}\right)\,,\\
\tilde{\sigma}_{xy}^{6r}\left(\epsilon\right)= & -\frac{1}{V\hbar}\sum_{\boldsymbol{k},i=1}^{6r}\Omega_{\bm{k}i}^{z}\delta\left(\epsilon-\hbar\omega_{\bm{k}i}\right)\,.
\end{align}
So the real energy magnetization $M_{E}^{z}$ is: 
\begin{multline}
M_{E}^{z}=-\frac{V}{4}\int d\epsilon\epsilon m_{1z}^{6r}\left(\epsilon\right)\frac{1}{\beta^{2}}\int^{\beta}d\beta^{\prime}\beta^{\prime}\left(2n\left(\epsilon\right)+\epsilon\frac{\partial n\left(\epsilon\right)}{\partial\epsilon}\right)\\
+\frac{V}{4}\int d\epsilon\epsilon^{3}\tilde{\sigma}_{xy}^{6r}\left(\epsilon\right)\frac{1}{\beta^{2}}\int^{\beta}d\beta^{\prime}\beta^{\prime}\frac{\partial n\left(\epsilon\right)}{\partial\epsilon}\,,\label{eq:energyM}
\end{multline}
where the boundary condition is that $2M_{E}^{z}$ coincides with
$\tilde{M}_{E}^{z}$ when $T=0$. There is no indefinity in Eq. \eqref{eq:energyM},
for the thermodynamic quantity~\cite{Qin2011} $\bm{M}_{s}=\bm{M}_{E}/T$
should be zero in the high temperature limit. We can change the integration
over $\beta^{\prime}$ into the integration over $\epsilon$ and we
have: 
\begin{align}
\kappa_{xy}^{\mathrm{tr}}= & \kappa_{xy}^{\mathrm{Kubo}}+\frac{2M_{E}^{z}}{TV}\,,\\
= & \frac{1}{2T}\int d\epsilon\tilde{\sigma}_{xy}^{6r}\left(\epsilon\right)\left(-2\int^{\epsilon}dxxn\left(x\right)+\epsilon^{2}n\left(\epsilon\right)\right)\,.
\end{align}
We can define $\tilde{\sigma}_{xy}^{6r}\left(\epsilon\right)=\frac{d\sigma_{xy}^{6r}\left(\epsilon\right)}{d\epsilon}$,
so we have:

\begin{multline}
\kappa_{xy}^{\mathrm{tr}}=\frac{1}{2T}\left.\sigma_{xy}^{6r}\left(\epsilon\right)\left(-2\int^{\epsilon}dxxn\left(x\right)+\epsilon^{2}n\left(\epsilon\right)\right)\right|_{-\infty}^{\infty}\\
-\frac{1}{2T}\int d\epsilon\sigma_{xy}^{6r}\left(\epsilon\right)\epsilon^{2}\frac{dn\left(\epsilon\right)}{d\epsilon}\,.\label{eq:partinte}
\end{multline}
The first term in Eq.\ \eqref{eq:partinte} is zero, for the following
reasons:

\begin{align}
\sigma_{xy}^{6r}\left(\epsilon\right) & =\int_{-\infty}^{\epsilon}dx\tilde{\sigma}_{xy}^{6r}\left(x\right)\,,\\
\sigma_{xy}^{6r}\left(\infty\right) & =0\,,\label{eq:infty}\\
\sigma_{xy}^{6r}\left(-\infty\right) & =0\,.\label{eq:-infty}
\end{align}
We can show Eq.\ \eqref{eq:infty} by the properties of the phonon
Berry curvature: $\Omega_{\bm{k}i}^{\left(-\right)}=-\Omega_{-\bm{k}i}^{\left(+\right)}$
where $1\leq i\leq3r$, because of the symmetry properties of $\psi_{\bm{k}i}$.
Eq.\ \eqref{eq:-infty} is zero because of the integration limit.
Therefore,

\begin{equation}
\kappa_{xy}^{\mathrm{tr}}=-\frac{1}{2T}\int d\epsilon\epsilon^{2}\sigma_{xy}^{6r}\left(\epsilon\right)\frac{dn\left(\epsilon\right)}{d\epsilon}\,.
\end{equation}
Next, we express $\kappa_{xy}^{\mathrm{tr}}$ with the $3r$ positive
energy bands. 
\begin{align}
\sigma_{xy}^{6r}\left(\epsilon\right) & =\int_{-\infty}^{0}dx\tilde{\sigma}_{xy}^{6r}\left(x\right)+\int_{0}^{\epsilon}dx\tilde{\sigma}_{xy}^{6r}\left(x\right)\,,\\
 & =-\frac{Z_{\mathrm{ph}}}{2\pi\hbar}+\int_{0}^{\epsilon}dx\tilde{\sigma}_{xy}^{6r}\left(x\right)\,.\label{eq:sigma}
\end{align}
Using our definition in Eq.\ \eqref{eq:phononconc}, we can write
Eq.\ \eqref{eq:sigma} as:

\begin{equation}
\sigma_{xy}^{6r}\left(\epsilon\right)=\begin{cases}
-\frac{Z_{\mathrm{ph}}}{2\pi\hbar}+\sigma_{xy}\left(\epsilon\right)\,, & \epsilon>0\\
-\frac{Z_{\mathrm{ph}}}{2\pi\hbar}+\sigma_{xy}\left(-\epsilon\right)\,. & \epsilon<0
\end{cases}
\end{equation}
Therefore, 
\begin{align}
\kappa_{xy}^{\mathrm{tr}}= & -\frac{1}{2T}\int_{0}^{\infty}d\epsilon\epsilon^{2}\left(-\frac{Z_{\mathrm{ph}}}{2\pi\hbar}+\sigma_{xy}\left(\epsilon\right)\right)\frac{dn\left(\epsilon\right)}{d\epsilon}\\
 & -\frac{1}{2T}\int_{-\infty}^{0}d\epsilon\epsilon^{2}\left(-\frac{Z_{\mathrm{ph}}}{2\pi\hbar}+\sigma_{xy}\left(-\epsilon\right)\right)\frac{dn\left(\epsilon\right)}{d\epsilon}.
\end{align}

Finally, after some simple algebra we obtain the phonon Hall conductivity
in Eq.~\eqref{eq:kappaxytr} and the topological term emerges naturally.

\section{TOPOLOGICAL PHONON SYSTEM}

\noindent The first term in Eq.~\eqref{eq:kappaxytr} is of the topological
nature, determined by the global phonon band structures. This is different
from the second term which is only determined by the low energy sectors
of the phonon bands limited by $k_{B}T$. For the two-dimensional
system (2D), $Z_{\mathrm{ph}}$ is the Chern number, quantized as
an integer. For the three-dimensional systems (3D), $Z_{\mathrm{ph}}$
is quantized~\cite{Halperin1987} in the unit $G_{z}/2\pi$, where
$G_{z}$ is the $z$-component of a reciprocal lattice lattice vector
$\bm{G}$. As the result, the phonon Hall conductivity has a topological
contribution with the quantized linear temperature coefficient in
the unit $(\pi k_{B})^{2}/3h$ (2D) or $\pi k_{B}^{2}G_{z}/6\pi h$
(3D).

Not surprisingly, most of the phonon systems have $Z_{\mathrm{ph}}=0$.
It is natural to define the topological phonon systems as those with:
\begin{equation}
Z_{\mathrm{ph}}\ne0\,.
\end{equation}
It puts a stringent constraint on what the real topological phonon
system is: it requires that the sum of the Chern numbers of all phonon
bands must be nonzero. The previous theoretical studies have successes
in constructing the phonon systems with non-zero Chern numbers of
individual bands~\cite{Prodan2009,Zhang2010,Zhang2011}. However,
these models rely on the reorganization of the phonon bands within
the positive energy branches, so still have zero $Z_{\mathrm{ph}}$.
They are not topological phonon systems in the stringent sense, nor
will they manifest unusual phonon Hall conductivity. To realize the
real topological phonon system, we need to look for insulating materials
with spin-orbit coupling so strong that the resulting effective magnetic
field can inter-mix and re-organize the phonon bands between the positive
and negative energy branches. A realistic lattice model for realizing
the topological phonon system is an interesting topic for future investigations.

We note that the value of the topological contribution to the thermal
Hall conductivity is actually the same as the longitudinal thermal
conductivity of a dielectric quantum wire with $Z_{\mathrm{ph}}$
acoustic phonon modes~\cite{Rego1998,Schwab2000,Yamamoto2004}. This
is because the topological phonon systems have chiral edge phonon
modes that behave just like the 1D acoustic phonon modes. The edge-bulk
correspondence picture in the topological phonon system is very similar
as that in quantum Hall systems. On the other hand, unlike the phonons
in a quantum wire, phonons in the chiral edge modes can not be backscattered
without coupling to the bulk. This will make topological phonon Hall
effect more robust against imperfections than the quantized thermal
conduction in 1D quantum wire. The latter had been observed experimentally~\cite{Schwab2000}.

\section{LOW TEMPERATURE LIMIT}

\noindent In the low temperature, only the low energy phonon modes
are relevant for the ordinary systems. It is thus sufficient to consider
the long wave acoustic phonon modes.

\subsection{Constraint for the effective magnetic field}

\noindent An important constraint for this case is: 
\begin{equation}
\sum_{l\kappa\kappa^{\prime}}G_{\alpha\beta}^{\kappa\kappa^{\prime}}(\bm{R}_{l}^{0})=0\,.\label{eq:G constraint}
\end{equation}
This can be verified directly by using the effective magnetic filed
in Eq.~\eqref{eq:GRealSpace}. As we will show, this is still true
even when we explicitly consider the external magnetic field because
of the overall charge neutrality. It is easy to see that: 
\begin{align}
 & \sum_{l\kappa\kappa^{\prime}}G_{\alpha\beta}^{\kappa\kappa^{\prime}}(\bm{R}_{l}^{0})\nonumber \\
= & \frac{2\hbar}{N}\mathrm{Im}\left.\left\langle \frac{\partial\Phi_{0}\left(\{\bm{R}^{0}+\bm{u}_{0}\}\right)}{\partial u_{0\beta}}\right|\left.\frac{\partial\Phi_{0}\left(\{\bm{R}^{0}+\bm{u}_{0}\}\right)}{\partial u_{0\alpha}}\right\rangle \right|_{\bm{u}_{0}\rightarrow0}\nonumber \\
 & +Z_{\mathrm{tot}}e\epsilon_{\alpha\beta\gamma}B_{\gamma}\,,\label{eq:sumG}
\end{align}
where $\Phi_{0}\left(\{\bm{R}^{0}+\bm{u}_{0}\}\right)$ is the ground
state wave-function of the electron subsystem when the whole system
is displaced with $\bm{u}_{0}$, and $Z_{\mathrm{tot}}$ is the total
charge number of the nuclei in an unit cell. We have assumed that
there is an external magnetic field $\bm{B}$. Since the displacement
is equivalent to a re-definition of the origin, we have: 
\begin{multline}
\Phi_{0}\left(\{\bm{R}^{0}+\bm{u}_{0}\},\{\bm{r}\}\right)=\exp\left[-\mathrm{i}\frac{e}{2\hbar}\left(\bm{B}\times\bm{u}_{0}\right)\cdot\left(\sum_{i}\bm{r}_{i}\right)\right]\\
\cdot\Phi_{0}\left(\{\bm{R}^{0}\},\{\bm{r}-\bm{u}_{0}\}\right)\,,\label{eq:Psi0shifted}
\end{multline}
where $\{\bm{r}\}\equiv\{\bm{r}_{1},\bm{r}_{2}\dots\bm{r}_{N_{e}}\}$
denotes the coordinates of the electron. We have: 
\begin{align}
 & \frac{\partial\Phi_{0}\left(\{\bm{R}^{0}+\bm{u}_{0}\},\{\bm{r}\}\right)}{\partial\bm{u}_{0}}\nonumber \\
= & -\frac{\mathrm{i}}{\hbar}\exp\left[-\mathrm{i}\frac{e}{2\hbar}\left(\bm{B}\times\bm{u}_{0}\right)\cdot\left(\sum_{i}\bm{r}_{i}\right)\right]\nonumber \\
 & \cdot\sum_{i}\left(\hat{\bm{p}}_{i}-\frac{e}{2}\bm{B}\times\bm{r}_{i}\right)\Phi_{0}\left(\{\bm{R}^{0}\},\{\bm{r}-\bm{u}_{0}\}\right)\,.\label{eq:pPisipu}
\end{align}

Substituting Eq.~(\ref{eq:pPisipu}) into (\ref{eq:sumG}), one can
easily verify that the first term and the second term of the right
hand side of Eq.~(\ref{eq:sumG}) exactly cancel each other due to
the overall charge neutrality.

\subsection{$T^{3}$ law }

\noindent It is easy to see that the phonon dynamics with the constraint
always has three acoustic modes that have zero energy at $\bm{k}=0$,
in consistent with the general requirement of global translational
symmetry. This remedies an important issue of the widely adopted phenomenological
model of Raman spin-lattice coupling, which has non-vanishing coupling
constant even in the long wave limit, inducing nonzero acoustic phonon
energies at $\bm{k}=0$. We will see that the constraint will change
the theoretical expectation of the low temperature behavior of the
phonon Hall conductivity.

With this constraint in mind, we can write down the general Hamiltonian
for the long-wave acoustic phonons of an isotropic continuous medium~\cite{Landau}:
\begin{equation}
\hat{H}=\int\mathrm{d}\bm{x}\left[\frac{\hat{\bm{P}}\left(\bm{x}\right)^{2}}{2\rho}+\frac{\mu_{1}}{2}\bm{\nabla}\hat{\bm{u}}\cdot\bm{\nabla}\hat{\bm{u}}+\frac{\mu_{2}}{2}\left(\bm{\nabla}\cdot\hat{\bm{u}}\right)^{2}\right]\,,\label{eq:continuousH}
\end{equation}
where $\hat{\bm{P}}\left(\bm{x}\right)=-\mathrm{i}\hbar\delta/\delta\hat{\bm{u}}\left(\bm{x}\right)-\bm{A}[\hat{\bm{u}}]$,
$\bm{u}\left(\bm{x}\right)$ is the vibration displacement, $\mu_{1}$
and $\mu_{2}$ are elastic constants, $\rho$ is the mass density.
The symmetry dictates: $\bm{A}[\hat{\bm{u}}]=\gamma_{1}\bm{\nabla}\bm{\nabla}\cdot\left(\bm{M}\times\hat{\bm{u}}\right)+\gamma_{2}\bm{\nabla}^{2}\left(\bm{M}\times\hat{\bm{u}}\right),$
where $\bm{M}$ is the magnetization of the system, and $\gamma_{1}$
and $\gamma_{2}$ are coupling constants characterizing the spin-lattice
coupling. With the Fourier transformation, we can identify $D_{\alpha\beta}(\bm{k})=(1/\rho)(\mu_{1}\delta_{\alpha\beta}k^{2}+\mu_{2}k_{\alpha}k_{\beta})$
and $G_{\alpha\beta}(\bm{k})=(1/\rho)\sum_{\gamma}\epsilon_{\alpha\beta\gamma}\left[-\gamma_{1}k_{\gamma}\bm{k}\cdot\bm{M}+\left(\gamma_{1}+2\gamma_{2}\right)\bm{k}^{2}M_{\gamma}\right]$
with $k=\left|\bm{k}\right|$ and $\alpha,\beta,\gamma=x,y,z$. We
can then adopt our general formula for calculating the phonon Hall
conductivity.

To the first order of the magnetization $\bm{M}$, we obtain the phonon
dispersions: $\omega_{\bm{k}1}=c_{L}k$ and $\omega_{\bm{k}2\left(3\right)}=c_{T}k\pm\gamma_{2}k\bm{k}\cdot\bm{M}$,
where $c_{T}=\sqrt{\mu_{1}/\rho}$ and $c_{L}=\sqrt{(\mu_{1}+2\mu_{2})/\rho}$
are transverse and longitudinal phonon velocities, respectively. The
phonon Berry curvatures are: 
\begin{align}
\bm{\Omega}_{\bm{k}1}= & -\frac{g_{1}\left(k^{2}\bm{M}+\bm{k}\bm{k}\cdot\bm{M}\right)}{k^{3}}\,,\label{eq:bcL}\\
\bm{\Omega}_{\bm{k}2\left(3\right)}= & \pm\frac{\bm{k}}{k^{3}}+\frac{g_{2}\left(k^{2}\bm{M}+\bm{k}\bm{k}\cdot\bm{M}\right)}{k^{3}}\,,\label{eq:bcT}
\end{align}
where $g_{1}=(\gamma_{1}+2\gamma_{2})(1+3\delta^{2})/(2c_{T}\delta(\delta^{2}-1))$,
$g_{2}=g_{1}\delta(3+\delta^{2})/(2(1+3\delta^{2}))$, and $\delta=c_{L}/c_{T}$.

Using Eq.~\eqref{eq:kappaxytr}, we determine the phonon Hall conductivity:
\begin{equation}
\kappa_{xy}^{\mathrm{tr}}=\frac{4\pi^{2}k_{B}^{4}}{45c_{T}^{3}\hbar^{3}}\left[1-\frac{\gamma_{1}+2\gamma_{2}}{2\gamma_{2}}\frac{4\delta^{3}+\delta^{2}+\delta+1}{\delta^{3}(\delta^{2}+\delta+1)}\right]\gamma_{2}M_{z}T^{3}\,,\label{eq:continuouskappa}
\end{equation}
where we assume $\bm{M}$ is along the $z$-direction and the Debye
energy $\hbar\omega_{D}\gg k_{B}T$. We can see that at the low temperature,
$\kappa_{xy}^{\mathrm{tr}}$ is proportional to $T^{3}$, instead
of $T$ as proposed in the previous studies~\cite{Sheng2006}.

We can obtain insights from the above calculation on how the phonon
would be deflected by the effective magnetic field. We can see from
Eqs. \eqref{eq:bcL} and \eqref{eq:bcT} that, on different branches,
the phonon will experience different ``reciprocal space magnetic
fields''. The corresponding anomalous velocity of the phonon is proportional
to $\bm{\Omega}_{\bm{k}i}\times\bm{\nabla}T$, similar to that for
electrons~\cite{Niu1999}. The net deflection direction will be perpendicular
to both the directions of the magnetization and the temperature gradient.

A few comments concerning the disorder effect~\cite{Kovalev2010}:
In analogue to the anomalous Hall effect of electron systems~\cite{Nagaosa2010},
we generally expect the total phonon Hall coefficient can be decomposed
to $\kappa_{xy}^{\mathrm{tr}}=\kappa_{xy}^{\mathrm{in}}+\kappa_{xy}^{\mathrm{sj}}+\kappa_{xy}^{\mathrm{skew}}$
, where $\kappa_{xy}^{\mathrm{in}}$ is the intrinsic phonon Hall
conductivity we calculate in Eq.~(\ref{eq:kappaxytr}), and the disorder
will introduce the side jump contribution $\kappa_{xy}^{\mathrm{sj}}$
and the skew scattering contribution $\kappa_{xy}^{\mathrm{skew}}$.
However, there is an important difference between the phonon system
and electron system: the mean free path of phonon in TGG had been
determined to be $\thicksim$1mm~\cite{Inyushkin2010}, much longer
than its electron counterpart. Moreover, in the low temperature limit,
the dominant contribution to the thermal conductivity is from the
long wave phonons which don't ``see'' the disorder~\cite{Flicker1973}.
We thus expect that these disorder correction is less important in
phonon systems, and the $T^{3}$ law of the phonon Hall conductivity
will survive.

\section{SUMMARY}

In summary, we establish the general phonon dynamics for the magnetic
solids. Based on the dynamics, we propose a general theory of the
PHE. Using the corrected Kubo formula, we link the intrinsic phonon
Hall conductivity to the phonon Berry curvature. The general formula
suggests that phonon could also have quantum Hall effect, and our
theory presents a rigorous definition of the topological phonon system.
We predict that the phonon Hall conductivity of the ordinary phonon
system is proportional to $T^{3}$ at the low temperature, while that
for the topological phonon system has the linear $T$ dependence with
the quantized temperature coefficient.

\section{APPENDIX}

\setcounter{equation}{0} \global\long\def\theequation{A\arabic{equation}}

\subsection{Phonon energy current operator and its scaling law}

\subsubsection{Phonon energy current operator\label{sub:Current}}

We follow Hardy\cite{Hardy1963} to derive the phonon energy current
operator. The Hamiltonian density is: 
\begin{equation}
\hat{h}\left(\boldsymbol{x}\right)=\frac{1}{2}\sum_{l\kappa}\left\{ \Delta\left(\boldsymbol{x}-\boldsymbol{R}_{l\kappa}\right)\hat{H}_{l\kappa}+\mathrm{h.c.}\right\} \,,
\end{equation}
where $\hat{H}_{l\kappa}=\frac{\hat{\boldsymbol{P}}_{l\kappa}^{2}}{2M_{\kappa}}+\hat{V}_{l\kappa}$
and $\hat{\bm{P}}_{l\kappa}=-\mathrm{i}\hbar\bm{\nabla}_{l\kappa}-\bm{A}_{l\kappa}$,
and $\Delta(\bm{x})$ is a localized function near $\bm{x}=0$ with
$\int\mathrm{d}\bm{x}\Delta(\bm{x})=1$. We adopt the Harmonic approximation
for $\hat{V}_{l\kappa}$: $\hat{V}_{l\kappa}=\frac{1}{2}\sum_{l^{\prime}\kappa^{\prime}\alpha\beta}\hat{u}_{l\kappa\alpha}D_{\alpha\beta}^{\kappa\kappa^{\prime}}\left(\bm{R}_{l}^{0}-\bm{R}_{l^{\prime}}^{0}\right)\hat{u}_{l^{\prime}\kappa^{\prime}\beta}$
where $D_{\alpha\beta}^{\kappa\kappa^{\prime}}\left(\bm{R}_{l}^{0}-\bm{R}_{l^{\prime}}^{0}\right)$
is the dynamical matrix. The Hamiltonian is $\hat{H}=\int\mathrm{d}\bm{x}\hat{h}(\bm{x})=\sum_{l\kappa}H_{l\kappa}$.
The energy current operator is defined by the energy conservation
equation:

\begin{equation}
\dot{\hat{h}}\left(\boldsymbol{x}\right)+\bm{\nabla}\cdot\hat{\boldsymbol{J}}_{E}\left(\boldsymbol{x}\right)=0\,,
\end{equation}
where $\hat{\boldsymbol{J}}_{E}\left(\boldsymbol{x}\right)$ is the
phonon energy current operator, and $\dot{\hat{h}}\left(\boldsymbol{x}\right)\equiv(1/\mathrm{i}\hbar)\left[\hat{h}\left(\boldsymbol{x}\right),\,\hat{H}\right]$.

We can express $\left[\hat{h}\left(\boldsymbol{x}\right),\,\hat{H}\right]$
as a divergence. We have:

\begin{align}
\frac{\mathrm{i}}{\hbar}\left[\hat{h}\left(\boldsymbol{x}\right),\,\hat{H}\right] & =\frac{\mathrm{i}}{2\hbar}\sum_{l\kappa,l^{\prime}\kappa^{\prime}}\Delta\left(\boldsymbol{x}-\boldsymbol{R}_{l\kappa}\right)\left[\hat{H}_{l\kappa},\,\hat{H}_{l^{\prime}\kappa^{\prime}}\right]\nonumber \\
 & +\mathrm{h.c.}+\cdots\,,\label{eq:hH}
\end{align}
where ``$\cdots$'' denotes the higher order terms such as $\left[\Delta\left(\boldsymbol{x}-\boldsymbol{R}_{l\kappa}\right),\,\frac{\hat{\boldsymbol{P}}_{l^{\prime}\kappa^{\prime}}^{2}}{2M_{\kappa^{\prime}}}\right]\hat{H}_{l\kappa}$,
which will be a cubic product of $\hat{\bm{u}}_{l\kappa}$ and $\hat{\bm{P}}_{l^{\prime}\kappa^{\prime}}$.
Noting that,

\begin{multline}
\left[\hat{H}_{l\kappa},\,\hat{H}_{l^{\prime}\kappa^{\prime}}\right]=\left[\frac{\hat{\boldsymbol{P}}_{l\kappa}^{2}}{2M_{\kappa}},\,\hat{V}_{l^{\prime}\kappa^{\prime}}\right]+\left[\hat{V}_{l\kappa},\,\frac{\hat{\boldsymbol{P}}_{l^{\prime}\kappa^{\prime}}^{2}}{2M_{\kappa^{\prime}}}\right]\\
+\frac{\mathrm{i}\hbar}{2M_{\kappa}M_{\kappa^{\prime}}}\sum_{\alpha\beta}G_{\alpha\beta}^{\kappa\kappa^{\prime}}\left(\bm{R}_{l}^{0}-\bm{R}_{l^{\prime}}^{0}\right)\left(\hat{P}_{l\kappa}^{\alpha}\hat{P}_{l^{\prime}\kappa^{\prime}}^{\beta}+\hat{P}_{l^{\prime}\kappa^{\prime}}^{\beta}\hat{P}_{l\kappa}^{\alpha}\right)\,,
\end{multline}
and interchanging subscripts $l\kappa$ and $l^{\prime}\kappa^{\prime}$,
we have:

\begin{align}
 & \frac{\mathrm{i}}{\hbar}\left[\hat{h}\left(\boldsymbol{x}\right),\,\hat{H}\right]\nonumber \\
= & \frac{\mathrm{i}}{2\hbar}\sum_{l\kappa,l^{\prime}\kappa^{\prime}}\left[\Delta\left(\boldsymbol{x}-\boldsymbol{R}_{l\kappa}\right)-\Delta\left(\bm{x}-\bm{R}_{l^{\prime}\kappa^{\prime}}\right)\right]\Biggl(\left[\frac{\hat{\boldsymbol{P}}_{l\kappa}^{2}}{2M_{\kappa}},\,\hat{V}_{l^{\prime}\kappa^{\prime}}\right]\nonumber \\
 & +\frac{\mathrm{i}\hbar}{2M_{\kappa}M_{\kappa^{\prime}}}\sum_{\alpha\beta}G_{\alpha\beta}^{\kappa\kappa^{\prime}}\left(\bm{R}_{l}^{0}-\bm{R}_{l^{\prime}}^{0}\right)\hat{P}_{l\kappa}^{\alpha}\hat{P}_{l^{\prime}\kappa^{\prime}}^{\beta}\Biggr)+\mathrm{h.c.}\,.\label{eq:harmonic}
\end{align}
Inserting the expansion 
\begin{multline}
\Delta\left(\boldsymbol{x}-\boldsymbol{R}_{l\kappa}\right)-\Delta\left(\bm{x}-\bm{R}_{l^{\prime}\kappa^{\prime}}\right)\approx\frac{1}{2}\left(\bm{R}_{l^{\prime}\kappa^{\prime}}^{0}-\bm{R}_{l\kappa}^{0}\right)\\
\cdot\left(\frac{\partial\Delta(\boldsymbol{x}-\boldsymbol{R}_{l^{\prime}\kappa^{\prime}}^{0})}{\partial\bm{x}}+\frac{\partial\Delta(\boldsymbol{x}-\boldsymbol{R}_{l\kappa}^{0})}{\partial\bm{x}}\right)\,,\label{eq:DeltaExpansion}
\end{multline}
into Eq.~\eqref{eq:harmonic}, we obtain: 
\begin{equation}
\frac{\mathrm{i}}{\hbar}\left[\hat{h}\left(\boldsymbol{x}\right),\,\hat{H}\right]=\bm{\nabla}\cdot\hat{\bm{J}}_{E}\left(\bm{x}\right)\,,\label{eq:PhononC}
\end{equation}
where

\begin{multline}
\hat{\bm{J}}_{E}\left(\bm{x}\right)=\frac{\mathrm{i}}{4\hbar}\sum_{l\kappa,l^{\prime}\kappa^{\prime}}\left(\bm{R}_{l^{\prime}\kappa^{\prime}}^{0}-\bm{R}_{l\kappa}^{0}\right)\left(\Delta\left(\boldsymbol{x}-\boldsymbol{R}_{l^{\prime}\kappa^{\prime}}^{0}\right)\right.\\
\left.+\Delta\left(\boldsymbol{x}-\boldsymbol{R}_{l\kappa}^{0}\right)\right)\Biggl(\left[\frac{\hat{\boldsymbol{P}}_{l\kappa}^{2}}{2M_{\kappa}},\,\hat{V}_{l^{\prime}\kappa^{\prime}}\right]+\\
+\frac{\mathrm{i}\hbar}{2M_{\kappa}M_{\kappa^{\prime}}}\sum_{\alpha\beta}G_{\alpha\beta}^{\kappa\kappa^{\prime}}\left(\bm{R}_{l}^{0}-\bm{R}_{l^{\prime}}^{0}\right)\hat{P}_{l\kappa}^{\alpha}\hat{P}_{l^{\prime}\kappa^{\prime}}^{\beta}\Biggr)+\mathrm{h.c.}\,.
\end{multline}

Doing the Fourier transformation $\hat{\bm{J}}_{E,\bm{q}}=\int d\bm{x}\hat{\bm{J}}_{E}\left(\bm{x}\right)e^{-\mathrm{i}\bm{q}\cdot\bm{x}}$,
we have: 
\begin{multline}
\hat{\bm{J}}_{E,\bm{q}}=-\frac{\mathrm{i}}{8}\Delta_{\bm{q}}\sum_{\bm{k}}\left(\hat{\bm{P}}_{\bm{k}}^{\dagger}\bm{\nabla}_{\bm{k}}\left(D_{\bm{k}}+D_{\bm{k}+\bm{q}}\right)\hat{\bm{u}}_{\bm{k}+\bm{q}}\right.\\
\left.-\hat{\bm{P}}_{\bm{k}}^{\dagger}\bm{\nabla}_{\bm{k}}\left(G_{\bm{k}}+G_{\bm{k}+\bm{q}}\right)\hat{\bm{P}}_{\bm{k}+\bm{q}}\right)\,,
\end{multline}
where $D_{\bm{k}}$ is the dynamic matrix with components $D_{\kappa\alpha,\kappa^{\prime}\beta}\left(\bm{k}\right)=\frac{1}{\sqrt{M_{\kappa}M_{\kappa^{\prime}}}}\sum_{l}D_{\alpha\beta}^{\kappa\kappa^{\prime}}(R_{l}^{0})e^{-\mathrm{i}\bm{k}\cdot(\bm{R}_{l}^{0}+\bm{d}_{\kappa\kappa^{\prime}})}$.
In the small $\bm{q}$ limit, $\Delta_{\bm{q}}\rightarrow1$. We obtain:

\begin{multline}
\hat{\bm{J}}_{E,\bm{q}}=\frac{1}{8}\sum_{\bm{k}}\hat{\Psi}_{\bm{k}}^{\dagger}\left(\bm{\nabla}_{\bm{k}}\left(\tilde{H}_{\bm{k}}+\tilde{H}_{\bm{k}+\bm{q}}\right)\right.\\
\left.+\bm{\nabla}_{\bm{k}}\left(\tilde{H}_{\bm{k}}^{\dagger}+\tilde{H}_{\bm{k}+\bm{q}}^{\dagger}\right)\right)\hat{\Psi}_{\bm{k}+\bm{q}}\,.
\end{multline}
Using the identities: $\bm{\nabla}_{\bm{k}}\tilde{H}_{\bm{k}}=\tilde{D}_{\bm{k}}\bm{\nabla}_{\bm{k}}\tilde{H}_{\bm{k}}$
and $\bm{\nabla}_{\bm{k}}\tilde{H}_{\bm{k}}^{\dagger}=\left(\bm{\nabla}_{\bm{k}}\tilde{H}_{\bm{k}}^{\dagger}\right)\tilde{D}_{\bm{k}}$,
we obtain Eq.~\eqref{eq:current}.

\begin{widetext}

\subsubsection{Scaling law for the energy current operator\label{sub:Scalinglaw}}

In order to calculate the energy magnetization, we need to verify
that the energy current satisfies the scaling law~\cite{Qin2011}:
\begin{equation}
\hat{\boldsymbol{J}}_{E}^{\psi}\left(\boldsymbol{x}\right)=\left(1+\psi\left(\bm{x}\right)\right)^{2}\hat{\bm{J}}_{E}\left(\bm{x}\right)+\mathcal{O}\left[\bm{\nabla}^{(2)}\psi(\bm{x})\right]\,,\label{eq:scaling}
\end{equation}
in the presence of the gravitational field $\psi\left(\bm{r}\right)$
that modifies the local Hamiltonian density by: 
\begin{equation}
\hat{h}^{\psi}\left(\boldsymbol{x}\right)=\frac{1}{2}\sum_{l\kappa}\left\{ \left(1+\psi\left(\bm{x}\right)\right)\Delta\left(\boldsymbol{x}-\boldsymbol{R}_{l\kappa}\right)\hat{H}_{l\kappa}+H.c.\right\} \,.
\end{equation}

Similar to the derivation in Sec.~\ref{sub:Current}, we have:

\begin{equation}
\bm{\nabla}\cdot\hat{\boldsymbol{J}}_{E}^{\psi}\left(\boldsymbol{x}\right)=\frac{\mathrm{i}}{\hbar}\left[\hat{h}^{\psi}\left(\boldsymbol{x}\right),\,\hat{H}^{\psi}\right]\,,\label{eq:motion}
\end{equation}
and: 
\begin{align}
\frac{\mathrm{i}}{\hbar}\left[\hat{h}^{\psi}\left(\boldsymbol{x}\right),\,\hat{H}^{\psi}\right] & =\frac{\mathrm{i}}{4}\int d\bm{x}^{\prime}\left(1+\psi\left(\bm{x}\right)\right)\left(1+\psi\left(\bm{x}^{\prime}\right)\right)\Gamma\left(\bm{x},\bm{x}^{\prime}\right)\,,\\
 & \approx\frac{\mathrm{i}}{4}\int d\bm{x}^{\prime}\left[\left(1+\psi\left(\bm{x}\right)\right)^{2}\Gamma\left(\bm{x},\bm{x}^{\prime}\right)+\left(1+\psi\left(\bm{x}\right)\right)\left(\bm{x}^{\prime}-\bm{x}\right)\cdot\frac{\partial\psi\left(\bm{x}\right)}{\partial\bm{x}}\Gamma\left(\bm{x},\bm{x}^{\prime}\right)\right.\label{eq:deltapsi1}\\
 & \left.+\frac{1}{2}\left(1+\psi\left(\bm{x}\right)\right)\sum_{\mu\nu}\left(x_{\mu}^{\prime}-x_{\mu}\right)\left(x_{\nu}^{\prime}-x_{\nu}\right)\frac{\partial^{2}\psi\left(\bm{x}\right)}{\partial x_{\mu}\partial x_{\nu}}\Gamma\left(\bm{x},\bm{x}^{\prime}\right)+\mathcal{O}\left[\bm{\nabla}^{(3)}\psi(\bm{x})\right]\right]\,.\label{eq:deltapsi2}
\end{align}
with 
\begin{equation}
\Gamma\left(\bm{x},\bm{x}^{\prime}\right)=\frac{1}{\hbar}\sum_{l\kappa,l^{\prime}\kappa^{\prime}}\Delta\left(\boldsymbol{x}-\boldsymbol{R}_{l\kappa}\right)\left\{ \Delta\left(\boldsymbol{x}^{\prime}-\boldsymbol{R}_{l^{\prime}\kappa^{\prime}}\right),\,\left[\hat{H}_{l\kappa},\,\hat{H}_{l^{\prime}\kappa^{\prime}}\right]\right\} +\mathrm{h.c.}+\cdots\,.\label{eq:Gamma}
\end{equation}
In Eq.\ \eqref{eq:Gamma}, we also ignore the terms which will lead
to the cubic products of $\hat{\bm{u}}_{l\kappa}$ and $\hat{\bm{P}}_{l^{\prime}\kappa^{\prime}}$
in the energy current operator. For the first term in Eq.~\eqref{eq:deltapsi1},
after the integration over $\bm{x}^{\prime}$ and repeating the derivation
from Eq.~\eqref{eq:hH} to Eq.~\eqref{eq:PhononC}, we can show:
\begin{equation}
\frac{\mathrm{i}}{4}\int d\bm{x}^{\prime}\left(1+\psi\left(\bm{x}\right)\right)^{2}\Gamma\left(\bm{x},\bm{x}^{\prime}\right)=\left(1+\psi\left(\bm{x}\right)\right)^{2}\bm{\nabla}\cdot\hat{\bm{J}}_{E}\left(\bm{x}\right)\,.\label{eq:(1+psi)2}
\end{equation}

For the second term in Eq.~\eqref{eq:deltapsi1}: 
\begin{align}
 & \frac{\mathrm{i}}{4}\int d\bm{x}^{\prime}\left(1+\psi\left(\bm{x}\right)\right)\left(\bm{x}^{\prime}-\bm{x}\right)\cdot\frac{\partial\psi\left(\bm{x}\right)}{\partial\bm{x}}\Gamma\left(\bm{x},\bm{x}^{\prime}\right)\\
= & \frac{\mathrm{i}}{4\hbar}\left(1+\psi\left(\bm{x}\right)\right)\frac{\partial\psi\left(\bm{x}\right)}{\partial\bm{x}}\cdot\sum_{l\kappa,l^{\prime}\kappa^{\prime}}\Delta\left(\boldsymbol{x}-\boldsymbol{R}_{l\kappa}\right)\left\{ \bm{R}_{l^{\prime}\kappa^{\prime}}^{0}-\bm{R}_{l\kappa}^{0},\,\left[\hat{H}_{l\kappa},\,\hat{H}_{l^{\prime}\kappa^{\prime}}\right]\right\} \,,\\
= & 2\left(1+\psi\left(\bm{x}\right)\right)\frac{\partial\psi\left(\bm{x}\right)}{\partial\bm{x}}\cdot\hat{\bm{J}}_{E}\left(\bm{x}\right)\,.\label{eq:nablapsi}
\end{align}

For the third term in Eq.~\eqref{eq:deltapsi2}, we have: 
\begin{align}
 & \frac{\mathrm{i}}{4}\int d\bm{x}^{\prime}\left(1+\psi\left(\bm{x}\right)\right)\sum_{\mu\nu}\left(x_{\mu}^{\prime}-x_{\mu}\right)\left(x_{\nu}^{\prime}-x_{\nu}\right)\frac{\partial^{2}\psi\left(\bm{x}\right)}{\partial x_{\mu}\partial x_{\nu}}\Gamma\left(\bm{x},\bm{x}^{\prime}\right)\\
= & \frac{\mathrm{i}}{2\hbar}\left(1+\psi\left(\bm{x}\right)\right)\sum_{l\kappa,l^{\prime}\kappa^{\prime},\mu\nu}\frac{\partial^{2}\psi\left(\bm{x}\right)}{\partial x_{\mu}\partial x_{\nu}}\Delta\left(\boldsymbol{x}-\boldsymbol{R}_{l\kappa}\right)\left(R_{l^{\prime}\kappa^{\prime}\mu}^{0}-R_{l\kappa\mu}^{0}\right)\left(R_{l^{\prime}\kappa^{\prime}\nu}^{0}-R_{l\kappa\nu}^{0}\right)\left[\hat{H}_{l\kappa},\,\hat{H}_{l^{\prime}\kappa^{\prime}}\right]+\mathrm{h.c.}\,,\label{eq:twoR}\\
= & \frac{\mathrm{i}}{4\hbar}\left(1+\psi\left(\bm{x}\right)\right)\sum_{l\kappa,l^{\prime}\kappa^{\prime},\mu\nu}\frac{\partial^{2}\psi\left(\bm{x}\right)}{\partial x_{\mu}\partial x_{\nu}}\left[\Delta\left(\boldsymbol{x}-\boldsymbol{R}_{l\kappa}\right)-\Delta\left(\boldsymbol{x}-\boldsymbol{R}_{l^{\prime}\kappa^{\prime}}\right)\right]\left(R_{l^{\prime}\kappa^{\prime}\mu}^{0}-R_{l\kappa\mu}^{0}\right)\left(R_{l^{\prime}\kappa^{\prime}\nu}^{0}-R_{l\kappa\nu}^{0}\right)\left[\hat{H}_{l\kappa},\,\hat{H}_{l^{\prime}\kappa^{\prime}}\right]+\mathrm{h.c.}\,.\label{eq:threeR}
\end{align}
\end{widetext}Noting the expansion in Eq.\ \eqref{eq:DeltaExpansion},
Eq.~\eqref{eq:threeR} doesn't have a contribution to the current
at the long wave limit.

Combining Eqs.~(\ref{eq:(1+psi)2}--\ref{eq:nablapsi}), we obtain
Eq.\ \eqref{eq:scaling}.

\subsection{Derivation details for $\kappa_{xy}^{\mathrm{Kubo}}$ in Eq.~\eqref{eq:kuboxy}}

Direct calculation of the Kubo formula leads to: 
\begin{equation}
\kappa_{xy}^{\mathrm{Kubo}}=\frac{\hbar}{32VT}\sum_{\bm{k};i,j=1}^{6r}\frac{\mathrm{Im}\left(\mathcal{V}_{\boldsymbol{k}ij}^{x}\mathcal{V}_{\bm{k}ji}^{y}-\mathcal{V}_{\bm{k}ij}^{y}\mathcal{V}_{\bm{k}ji}^{x}\right)\omega_{\bm{k}i}\omega_{\bm{k}j}n_{\bm{k}i}}{\left(\omega_{\bm{k}i}-\omega_{\bm{k}j}\right)^{2}}\,,\label{eq:vv}
\end{equation}
where $\bm{\mathcal{V}}_{\bm{k}ij}\equiv2\left(\bar{\psi}_{\bm{k}i}\frac{\partial\tilde{H}_{\bm{k}}}{\partial\bm{k}}\psi_{\bm{k}j}+\psi_{\bm{k}i}^{\dagger}\frac{\partial\tilde{H}_{\bm{k}}^{\dagger}}{\partial\bm{k}}\bar{\psi}_{\bm{k}j}^{\dagger}\right)$.

It is easy to verify: 
\begin{align}
\bar{\psi}_{\bm{k}i}\frac{\partial\tilde{H}_{\bm{k}}}{\partial k_{x}}\psi_{\bm{k}j} & =\left(\omega_{\bm{k}j}-\omega_{\bm{k}i}\right)\bar{\psi}_{\bm{k}i}\frac{\partial\psi_{\bm{k}j}}{\partial k_{x}}+\frac{\partial\omega_{\bm{k}i}}{\partial k_{x}}\delta_{ij}\,,
\end{align}
so: 
\begin{align}
\mathcal{V}_{\bm{k}ij}^{x}= & 4\frac{\partial\omega_{\bm{k}i}}{\partial k_{x}}\delta_{ij}+2\left(\omega_{\bm{k}j}-\omega_{\bm{k}i}\right)\left(\bar{\psi}_{\bm{k}i}\frac{\partial\psi_{\bm{k}j}}{\partial k_{x}}+\psi_{\bm{k}i}^{\dagger}\frac{\partial\bar{\psi}_{\bm{k}j}^{\dagger}}{\partial k_{x}}\right)\,.\label{eq:V}
\end{align}
Note here $i\neq j$, so we have:

\begin{multline}
\sum_{j=1}^{6r}\frac{\mathcal{V}_{\bm{k}ij}^{x}\mathcal{V}_{\bm{k}ji}^{y}\omega_{\bm{k}i}\omega_{\bm{k}j}}{\left(\omega_{\bm{k}i}-\omega_{\bm{k}j}\right)^{2}}=4\omega_{\bm{k}i}\left(\frac{\partial\bar{\psi}_{\bm{k}i}}{\partial k_{x}}\tilde{H}_{\bm{k}}\frac{\partial\psi_{\bm{k}i}}{\partial k_{y}}\right.\\
+\frac{\partial\psi_{\bm{k}i}^{\dagger}}{\partial k_{x}}\tilde{H}_{\bm{k}}^{\dagger}\frac{\partial\bar{\psi}_{\bm{k}i}^{\dagger}}{\partial k_{y}}+\frac{\partial\psi_{\bm{k}i}^{\dagger}}{\partial k_{x}}\tilde{H}_{\bm{k}}^{\dagger}\tilde{D}_{\bm{k}}\frac{\partial\psi_{\bm{k}i}}{\partial k_{y}}\\
\left.+\frac{\partial\bar{\psi}_{\bm{k}i}}{\partial k_{x}}\tilde{H}_{\bm{k}}\tilde{D}_{\bm{k}}^{-1}\frac{\partial\bar{\psi}_{\bm{k}i}^{\dagger}}{\partial k_{y}}\right)\,.\label{eq:gammaxy}
\end{multline}
where we have used $\bar{\psi}_{\bm{k}i}\frac{\partial\psi_{\bm{k}j}}{\partial k_{x}}=-\frac{\partial\bar{\psi}_{\bm{k}i}}{\partial k_{x}}\psi_{\bm{k}j}$,
$\psi_{\bm{k}j}^{\dagger}\frac{\partial\bar{\psi}_{\bm{k}i}^{\dagger}}{\partial k_{x}}=-\frac{\partial\psi_{\bm{k}j}^{\dagger}}{\partial k_{x}}\bar{\psi}_{\bm{k}i}^{\dagger}$,
and $\tilde{H}_{\bm{k}}=\sum_{j}\omega_{\bm{k}j}\psi_{\bm{k}j}\bar{\psi}_{\bm{k}j}$
and its hermitian conjugate. We can further simplify the last two
terms in Eq.~\eqref{eq:gammaxy} using $\tilde{D}_{\bm{k}}\frac{\partial\psi_{\bm{k}i}}{\partial k_{y}}=\frac{\partial\bar{\psi}_{\bm{k}i}^{\dagger}}{\partial k_{y}}-\frac{\partial\tilde{D}_{\bm{k}}}{\partial k_{y}}\psi_{\bm{k}i}$
and $\tilde{D}_{\bm{k}}^{-1}\frac{\partial\bar{\psi}_{\bm{k}i}^{\dagger}}{\partial k_{y}}=\frac{\partial\psi_{\bm{k}i}}{\partial k_{y}}-\frac{\partial\tilde{D}_{\bm{k}}^{-1}}{\partial k_{y}}\bar{\psi}_{\bm{k}i}^{\dagger}$,
and we obtain:

\begin{align}
 & \frac{\partial\psi_{\bm{k}i}^{\dagger}}{\partial k_{x}}\tilde{H}_{\bm{k}}^{\dagger}\tilde{D}_{\bm{k}}\frac{\partial\psi_{\bm{k}i}}{\partial k_{y}}+\frac{\partial\bar{\psi}_{\bm{k}i}}{\partial k_{x}}\tilde{H}_{\bm{k}}\tilde{D}_{\bm{k}}^{-1}\frac{\partial\bar{\psi}_{\bm{k}i}^{\dagger}}{\partial k_{y}}\label{eq:nonzero}\\
= & \frac{\partial\psi_{\bm{k}i}^{\dagger}}{\partial k_{x}}\tilde{H}_{\bm{k}}^{\dagger}\frac{\partial\bar{\psi}_{\bm{k}i}^{\dagger}}{\partial k_{y}}+\frac{\partial\bar{\psi}_{\bm{k}i}}{\partial k_{x}}\tilde{H}_{\bm{k}}\frac{\partial\psi_{\bm{k}i}}{\partial k_{y}}\nonumber \\
- & \frac{\partial\psi_{\bm{k}i}^{\dagger}}{\partial k_{x}}\tilde{H}_{\bm{k}}^{\dagger}\frac{\partial\tilde{D}_{\bm{k}}}{\partial k_{y}}\psi_{\bm{k}i}-\frac{\partial\bar{\psi}_{\bm{k}i}}{\partial k_{x}}\tilde{H}_{\bm{k}}\frac{\partial\tilde{D}_{\bm{k}}^{-1}}{\partial k_{y}}\bar{\psi}_{\bm{k}i}^{\dagger}\,.\label{eq:zero}
\end{align}
The second term in Eq.~\eqref{eq:zero}: $\frac{\partial\bar{\psi}_{\bm{k}i}}{\partial k_{x}}\tilde{H}_{\bm{k}}\frac{\partial\tilde{D}_{\bm{k}}^{-1}}{\partial k_{y}}\bar{\psi}_{\bm{k}i}^{\dagger}=\psi_{\bm{k}i}^{\dagger}\frac{\partial D_{\bm{k}}}{\partial k_{x}}\tilde{H}_{\bm{k}}\frac{\partial\tilde{D}_{\bm{k}}^{-1}}{\partial k_{y}}\bar{\psi}_{\bm{k}i}^{\dagger}-\frac{\partial\psi_{\bm{k}i}^{\dagger}}{\partial k_{x}}\tilde{D}_{\bm{k}}\tilde{H}_{\bm{k}}\tilde{D}_{\bm{k}}^{-1}\frac{\partial\tilde{D}_{\bm{k}}}{\partial k_{y}}\psi_{\bm{k}i}=-\frac{\partial\psi_{\bm{k}i}^{\dagger}}{\partial k_{x}}\tilde{H}_{\bm{k}}^{\dagger}\frac{\partial\tilde{D}_{\bm{k}}}{\partial k_{y}}\psi_{\bm{k}i}$,
and $\frac{\partial\tilde{D}_{\bm{k}}}{\partial k_{x}}\tilde{H}_{\bm{k}}\frac{\partial\tilde{D}_{\bm{k}}^{-1}}{\partial k_{y}}=0$
for $\frac{\partial\tilde{D}_{\bm{k}}}{\partial k_{x}}=\left[\begin{array}{cc}
\frac{\partial D_{\bm{k}}}{\partial k_{x}} & 0\\
0 & 0
\end{array}\right]$, and then the two terms in Eq.~\eqref{eq:zero} vanish. Substituting
Eq.~\eqref{eq:nonzero} into Eq.~\eqref{eq:gammaxy}, we have:

\begin{align}
 & \sum_{j=1}^{6r}\frac{\mathrm{Im}\left(\mathcal{V}_{\bm{k}ij}^{x}\mathcal{V}_{\bm{k}ji}^{y}\right)\omega_{\bm{k}i}\omega_{\bm{k}j}}{\left(\omega_{\bm{k}i}-\omega_{\bm{k}j}\right)^{2}}\nonumber \\
= & 8\omega_{\bm{k}i}\mathrm{Im}\left(\frac{\partial\bar{\psi}_{\bm{k}i}}{\partial k_{x}}\tilde{H}_{\bm{k}}\frac{\partial\psi_{\bm{k}i}}{\partial k_{y}}-\frac{\partial\bar{\psi}_{\bm{k}i}}{\partial k_{y}}\tilde{H}_{\bm{k}}\frac{\partial\psi_{\bm{k}i}}{\partial k_{x}}\right)\,.
\end{align}
Finally, we obtain:

\begin{align}
\kappa_{xy}^{\mathrm{Kubo}}= & \frac{\hbar}{2VT}\mathrm{Im}\sum_{\bm{k},i=1}^{6r}\left(\frac{\partial\bar{\psi}_{\bm{k}i}}{\partial k_{x}}\tilde{H}_{\bm{k}}\frac{\partial\psi_{\bm{k}i}}{\partial k_{y}}-\frac{\partial\bar{\psi}_{\bm{k}i}}{\partial k_{y}}\tilde{H}_{\bm{k}}\frac{\partial\psi_{\bm{k}i}}{\partial k_{x}}\right)\nonumber \\
 & \cdot\omega_{\bm{k}i}n_{\bm{k}i}\,,\label{eq:6r}\\
= & \frac{\hbar}{2VT}\mathrm{Im}\sum_{\bm{k},i=1}^{3r}\left(\frac{\partial\bar{\psi}_{\bm{k}i}}{\partial k_{x}}\tilde{H}_{\bm{k}}\frac{\partial\psi_{\bm{k}i}}{\partial k_{y}}-\frac{\partial\bar{\psi}_{\bm{k}i}}{\partial k_{y}}\tilde{H}_{\bm{k}}\frac{\partial\psi_{\bm{k}i}}{\partial k_{x}}\right)\nonumber \\
 & \cdot\omega_{\bm{k}i}\left(2n_{\bm{k}i}+1\right)\,.\label{eq:3r}
\end{align}
From Eq.~\eqref{eq:6r} to Eq.~\eqref{eq:3r}, we have used the
symmetry properties of $\omega_{\bm{k}i}$ and $\psi_{\bm{k}i}$,
so we come to Eq.~\eqref{eq:kuboxy}.

\subsection{Derivation details for $\tilde{M}_{E}^{z}$ in Eq.~\eqref{eq:PEM}}

After a direct calculation of the canonical correlation function in
Eq.~\eqref{eq:MEtilt}, we obtain: 
\begin{multline}
\tilde{M}_{E}^{z}=\frac{\mathrm{i}\hbar}{16}\frac{\partial}{\partial q_{y}}\sum_{\bm{k};i,j=1}^{6r}S_{\boldsymbol{k}+\bm{q},\boldsymbol{k}ji}\left(\mathcal{V}_{\boldsymbol{k},\bm{k}+\bm{q}ij}^{x}+\mathcal{V}_{\boldsymbol{k}+\bm{q},\bm{k}ji}^{x\ast}\right)\\
\left.\cdot\omega_{\bm{k}i}\omega_{\bm{k}+\bm{q}j}\frac{n_{\bm{k}+\bm{q}j}-n_{\bm{k}i}}{\omega_{\bm{k}i}-\omega_{\bm{k}+\bm{q}j}}\right|_{\bm{q}\rightarrow0}\,,\label{eq:tildeMz}
\end{multline}
where $S_{\bm{k}+\bm{q},\bm{k}ji}\equiv\bar{\psi}_{\bm{k}+\bm{q}j}\psi_{\bm{k}i}$
and $\bm{\mathcal{V}}_{\bm{k},\bm{k}+\bm{q}ij}\equiv\bar{\psi}_{\bm{k}i}\frac{\partial\left(\tilde{H}_{\bm{k}}+\tilde{H}_{\bm{k}+\bm{q}}\right)}{\partial\bm{k}}\psi_{\bm{k}+\bm{q}j}+\psi_{\bm{k}i}^{\dagger}\frac{\partial\left(\tilde{H}_{\bm{k}}^{\dagger}+\tilde{H}_{\bm{k}+\bm{q}}^{\dagger}\right)}{\partial\bm{k}}\bar{\psi}_{\bm{k}+\bm{q}j}^{\dagger}.$

First, we calculate the inter-band contribution from the terms with
$i\neq j$ in Eq.~\eqref{eq:tildeMz}. When $\bm{q}\rightarrow0$,
we have:

\begin{equation}
\tilde{M}_{E}^{z,\mathrm{inter}}=-\frac{\hbar}{8}\sum_{\bm{k};i,j=1}^{6r}\frac{\mathrm{Im}\left(\frac{\partial\bar{\psi}_{\bm{k}j}}{\partial k_{y}}\psi_{\bm{k}i}\mathcal{V}_{\boldsymbol{k}ij}^{x}\right)\omega_{\bm{k}i}\omega_{\bm{k}j}\left(n_{\bm{k}j}-n_{\bm{k}i}\right)}{\omega_{\bm{k}i}-\omega_{\bm{k}j}}\,.\label{eq:minter}
\end{equation}
We further have:

\begin{equation}
\mathrm{Im}\left(\frac{\partial\bar{\psi}_{\bm{k}j}}{\partial k_{y}}\psi_{\bm{k}i}\mathcal{V}_{\boldsymbol{k}ij}^{x}\right)=\frac{1}{2}\mathrm{Im}\left(\frac{\partial\bar{\psi}_{\bm{k}j}}{\partial k_{y}}\psi_{\bm{k}i}\mathcal{V}_{\boldsymbol{k}ij}^{x}-\psi_{\bm{k}i}^{\dagger}\frac{\partial\bar{\psi}_{\bm{k}j}^{\dagger}}{\partial k_{y}}\mathcal{V}_{\boldsymbol{k}ij}^{x\ast}\right)\,,\label{eq:SGamma}
\end{equation}
Noting $\mathcal{V}_{\bm{k}ij}^{\ast}=\mathcal{V}_{\bm{k}ji}$ and
inserting Eq.~\eqref{eq:SGamma} into Eq.~\eqref{eq:minter}, we
obtain:

\begin{multline}
\tilde{M}_{E}^{z,\mathrm{inter}}=-\frac{\hbar}{16}\sum_{\bm{k};i,j=1}^{6r}\mathrm{Im}\left[\left(\frac{\partial\bar{\psi}_{\bm{k}j}}{\partial k_{y}}\psi_{\bm{k}i}-\psi_{\bm{k}j}^{\dagger}\frac{\partial\bar{\psi}_{\bm{k}i}^{\dagger}}{\partial k_{y}}\right)\mathcal{V}_{\boldsymbol{k}ij}^{x}\right]\\
\cdot\frac{\omega_{\bm{k}i}\omega_{\bm{k}j}\left(n_{\bm{k}j}-n_{\bm{k}i}\right)}{\omega_{\bm{k}i}-\omega_{\bm{k}j}}\,,
\end{multline}
where we have interchanged $i$ and $j$ of the second term in Eq.~\eqref{eq:SGamma}.
Using Eq.~\eqref{eq:V}, we have 
\begin{multline}
\tilde{M}_{E}^{z,\mathrm{inter}}=\frac{\hbar}{32}\sum_{\bm{k};i,j=1}^{6r}\frac{\mathrm{Im}\left(\mathcal{V}_{\bm{k}ji}^{y}\mathcal{V}_{\boldsymbol{k}ij}^{x}\right)\omega_{\bm{k}i}\omega_{\bm{k}j}\left(n_{\bm{k}j}-n_{\bm{k}i}\right)}{\left(\omega_{\bm{k}i}-\omega_{\bm{k}j}\right)^{2}}\,,\\
=-\frac{\hbar}{32}\sum_{\bm{k};i,j=1}^{6r}\frac{\mathrm{Im}\left(\mathcal{V}_{\boldsymbol{k}ij}^{x}\mathcal{V}_{\bm{k}ji}^{y}-\mathcal{V}_{\bm{k}ij}^{y}\mathcal{V}_{\bm{k}ji}^{x}\right)\omega_{\bm{k}i}\omega_{\bm{k}j}n_{\bm{k}i}}{\left(\omega_{\bm{k}i}-\omega_{\bm{k}j}\right)^{2}}\,.\label{eq:MKappa}
\end{multline}
Comparing Eq.~\eqref{eq:MKappa} and~\eqref{eq:vv}, we finally
obtain: 
\begin{align}
\tilde{M}_{E}^{z,\mathrm{inter}}= & -\frac{\hbar}{2}\mathrm{Im}\sum_{\bm{k};i=1}^{3r}\left(\frac{\partial\bar{\psi}_{\bm{k}i}}{\partial k_{x}}\tilde{H}_{\bm{k}}\frac{\partial\psi_{\bm{k}i}}{\partial k_{y}}-\frac{\partial\bar{\psi}_{\bm{k}i}}{\partial k_{y}}\tilde{H}_{\bm{k}}\frac{\partial\psi_{\bm{k}i}}{\partial k_{x}}\right)\nonumber \\
 & \cdot\omega_{\bm{k}i}\left(2n_{\bm{k}i}+1\right)\\
= & -\frac{\hbar}{2}\sum_{\bm{k};i=1}^{3r}\mathcal{M}_{\bm{k}i}^{z}\omega_{\bm{k}i}\left(2n_{\bm{k}i}+1\right)\,.\label{eq:Minter}
\end{align}

Second, we calculate the intra-band contribution from the term with
$i=j$ in Eq.~\eqref{eq:tildeMz}:

\begin{multline}
\tilde{M}_{E}^{z,\mathrm{intra}}=-\frac{\hbar}{16}\frac{\partial}{\partial q_{y}}\sum_{\bm{k};i=1}^{6r}\mathrm{Im}\left[S_{\boldsymbol{k}+\bm{q},\boldsymbol{k}i}\left(\mathcal{V}_{\boldsymbol{k},\bm{k}+\bm{q}i}^{x}+\mathcal{V}_{\boldsymbol{k}+\bm{q},\bm{k}i}^{x\ast}\right)\right]\\
\cdot\left.\frac{\omega_{\bm{k}i}\omega_{\bm{k}+\bm{q}i}\left(n_{\bm{k}+\bm{q}i}-n_{\bm{k}i}\right)}{\omega_{\bm{k}i}-\omega_{\bm{k}+\bm{q}i}}\right|_{\bm{q}\rightarrow0}\,.
\end{multline}
For further simplification, note:

\[
\left.\frac{\partial S_{\boldsymbol{k}+\bm{q},\boldsymbol{k}i}}{\partial q_{y}}\right|_{\bm{q}\rightarrow0}=\frac{\partial\bar{\psi}_{\bm{k}i}}{\partial k_{y}}\psi_{\bm{k}i}\,,
\]

\begin{multline}
\left.\frac{\partial\left(\mathcal{V}_{\boldsymbol{k},\bm{k}+\bm{q}i}^{x}+\mathcal{V}_{\boldsymbol{k}+\bm{q},\bm{k}i}^{x\ast}\right)}{\partial q_{y}}\right|_{\bm{q}\rightarrow0}=2\left(2\bar{\psi}_{\bm{k}i}\frac{\partial\tilde{H}_{\bm{k}}}{\partial k_{x}}\frac{\partial\psi_{\bm{k}i}}{\partial k_{y}}\right.\\
\left.+2\psi_{\bm{k}i}^{\dagger}\frac{\partial\tilde{H}_{\bm{k}}^{\dagger}}{\partial k_{x}}\frac{\partial\bar{\psi}_{\bm{k}i}^{\dagger}}{\partial k_{y}}+\bar{\psi}_{\bm{k}i}\frac{\partial^{2}\tilde{H}_{\bm{k}}}{\partial k_{x}\partial k_{y}}\psi_{\bm{k}i}+\psi_{\bm{k}i}^{\dagger}\frac{\partial^{2}\tilde{H}_{\bm{k}}^{\dagger}}{\partial k_{x}\partial k_{y}}\bar{\psi}_{\bm{k}i}^{\dagger}\right)\,,\label{eq:Vqy}
\end{multline}

\[
\left.\frac{n_{\bm{k}+\bm{q}i}-n_{\bm{k}i}}{\omega_{\bm{k}+\bm{q}i}-\omega_{\bm{k}i}}\right|_{\bm{q}\rightarrow0}=n_{\bm{k}i}^{\prime}\,.
\]

One can show that the sum of the last two terms in Eq.~\eqref{eq:Vqy}
is real, so it has no contribution. So we can write:

\begin{multline}
\tilde{M}_{E}^{z,\mathrm{intra}}=\frac{\hbar}{4}\sum_{\bm{k};i=1}^{6r}\mathrm{Im}\left[2\frac{\partial\bar{\psi}_{\bm{k}i}}{\partial k_{y}}\psi_{\bm{k}i}\frac{\partial\omega_{\bm{k}i}}{\partial k_{x}}+\bar{\psi}_{\bm{k}i}\frac{\partial\tilde{H}_{\bm{k}}}{\partial k_{x}}\frac{\partial\psi_{\bm{k}i}}{\partial k_{y}}\right.\\
\left.+\psi_{\bm{k}i}^{\dagger}\frac{\partial\tilde{H}_{\bm{k}}^{\dagger}}{\partial k_{x}}\frac{\partial\bar{\psi}_{\bm{k}i}^{\dagger}}{\partial k_{y}}\right]\omega_{\bm{k}i}^{2}n_{\bm{k}i}^{\prime}\,,
\end{multline}
With $\mathrm{Im}[\frac{\partial\bar{\psi}_{\bm{k}i}}{\partial k_{y}}\psi_{\bm{k}i}]=-\mathrm{Im}[\bar{\psi}_{\bm{k}i}\frac{\partial\psi_{\bm{k}i}}{\partial k_{y}}]=\mathrm{Im}[\frac{\partial\psi_{\bm{k}i}^{\dagger}}{\partial k_{y}}\bar{\psi}_{\bm{k}i}^{\dagger}]$,
we come to 
\begin{multline}
\tilde{M}_{E}^{z,\mathrm{intra}}=\frac{\hbar}{4}\sum_{\bm{k};i=1}^{6r}\mathrm{Im}\left[\bar{\psi}_{\bm{k}i}\frac{\partial\left(\tilde{H}_{\bm{k}}-\omega_{\bm{k}i}\right)}{\partial k_{x}}\frac{\partial\psi_{\bm{k}i}}{\partial k_{y}}\right.\\
\left.+\psi_{\bm{k}i}^{\dagger}\frac{\partial\left(\tilde{H}_{\bm{k}}^{\dagger}-\omega_{\bm{k}i}\right)}{\partial k_{x}}\frac{\partial\bar{\psi}_{\bm{k}i}^{\dagger}}{\partial k_{y}}\right]\omega_{\bm{k}i}^{2}n_{\bm{k}i}^{\prime}\,.
\end{multline}
Using $\bar{\psi}_{\bm{k}i}\frac{\partial\left(\tilde{H}_{\bm{k}}-\omega_{\bm{k}i}\right)}{\partial k_{x}}=-\frac{\partial\bar{\psi}_{\bm{k}i}}{\partial k_{x}}\left(\tilde{H}_{\bm{k}}-\omega_{\bm{k}i}\right)$
and $\psi_{\bm{k}i}^{\dagger}\frac{\partial\left(\tilde{H}_{\bm{k}}^{\dagger}-\omega_{\bm{k}i}\right)}{\partial k_{x}}=-\frac{\partial\psi_{\bm{k}i}^{\dagger}}{\partial k_{x}}\left(\tilde{H}_{\bm{k}}^{\dagger}-\omega_{\bm{k}i}\right)$,
we obtain: 
\begin{align}
\tilde{M}_{E}^{z,\mathrm{intra}} & =-\frac{\hbar}{4}\sum_{\bm{k};i=1}^{6r}\mathrm{Im}\left[\frac{\partial\bar{\psi}_{\bm{k}i}}{\partial k_{x}}\left(\tilde{H}_{\bm{k}}-\omega_{\bm{k}i}\right)\frac{\partial\psi_{\bm{k}i}}{\partial k_{y}}\right.\nonumber \\
 & \left.-\frac{\partial\bar{\psi}_{\bm{k}i}}{\partial k_{y}}\left(\tilde{H}_{\bm{k}}-\omega_{\bm{k}i}\right)\frac{\partial\psi_{\bm{k}i}}{\partial k_{x}}\right]\omega_{\bm{k}i}^{2}n_{\bm{k}i}^{\prime}\,.
\end{align}
Using the definitions for $\mathcal{M}_{\bm{k}i}^{z}$ and $\Omega_{\bm{k}i}^{z}$,
we have: 
\begin{align}
\tilde{M}_{E}^{z,\mathrm{intra}} & =-\frac{\hbar}{4}\sum_{\bm{k};i=1}^{6r}\left(\mathcal{M}_{\bm{k}i}^{z}+\omega_{\bm{k}i}\Omega_{\bm{k}i}^{z}\right)\omega_{\bm{k}i}^{2}n_{\bm{k}i}^{\prime}\,,\\
 & =-\frac{\hbar}{2}\sum_{\bm{k};i=1}^{3r}\left(\mathcal{M}_{\bm{k}i}^{z}+\omega_{\bm{k}i}\Omega_{\bm{k}i}^{z}\right)\omega_{\bm{k}i}^{2}n_{\bm{k}i}^{\prime}\,.\label{eq:Mintra}
\end{align}
Finally, combining Eqs.~\eqref{eq:Minter} and~\eqref{eq:Mintra},
we come to Eq.~\eqref{eq:PEM}.

\end{document}